\documentclass[12pt]{iopart}

\usepackage{iopams}
\usepackage{amsmath2}
\usepackage{graphicx}

\newcommand{\RN}{Reissner-Nordstr\"{o}m }
\newcommand{\SC}{Schwarzschild }
\newcommand{\KS}{Kruskal-Szekeres }
\newcommand{\Rp}{r_{+}}
\newcommand{\Rm}{r_{-}}

\begin{document}

\title[Embedding the \RN spacetime]{Embedding the \RN spacetime in Euclidean and Minkowski spaces}
\author{Uri Jacob and Tsvi Piran}
\address{Racah Institute of Physics, Hebrew University, Jerusalem, Israel}
\eads{\mailto{uriyada@phys.huji.ac.il}, \mailto{tsvi@phys.huji.ac.il}}

\begin{abstract}
We examine embedding diagrams of hypersurfaces in the \RN black hole spacetime. These
embedding diagrams serve as useful tools to visualize the geometry of the hypersurfaces
and of the whole spacetime in general.
\end{abstract}


\section{Introduction}
The \RN (RN) spacetime describes a vacuum where matter and
electric charge exist in a singularity at the center $(r=0)$. When
the mass is large enough compared to the electric charge, $M>Q$,
which is the case we explore here, a black hole engulfs the
singularity. This black hole is far more complex than the simple
\SC black hole.
\par
The RN spacetime has been extensively investigated in the past
(see e.g. \cite{Gravitation,Chandrasekhar,Graves}). Our aim here
is to describe the embedding diagrams of typical hypersurfaces of
this spacetime. The embedding diagrams of spatial
hypersurfaces of the \SC black hole are well known (see
\cite{Gravitation}). Surprisingly, little is known about the
corresponding diagrams of the richer RN (and the rather similar
Kerr) spacetime. We review briefly the concept of embedding
diagrams in Sec. \ref{EMBsec} and discuss as an example the
embedding of the \SC spacetime. We present in Sec. \ref{COOsec} RN
coordinate systems that enable us to explore the RN spacetime, and
we derive the formulas to calculate the embeddings of RN
hypersurfaces. In Sec. \ref{RNEMBsec} we present the embedding
diagrams of these slices. We summarize our results in Sec.
\ref{SUMsec}.

\section{Embedding diagrams}\label{EMBsec}
Embedding diagrams of hypersurfaces of a spacetime are useful
tools to visualize the overall geometry. Consider a 3-dimensional
hypersurface with an induced metric of
$ds^2=\tilde{g}_{ij}dx^idx^j$. When the hypersurface is
spherically symmetric (and $r \neq const$ on the hypersurface),
we can select a 2-dimensional plane within
this hypersurface and express the metric as:
\begin{equation}
ds^2=\tilde{g}_{rr}dr^2+r^2d\varphi^2.
\end{equation}
This 2-dimensional plane can now be embedded in a flat
3-dimensional space:
\begin{equation}
ds^2=\pm dz^2+dr^2+r^2d\varphi^2 .
\end{equation}
This embedding space is Euclidean or Minkowski depending on
whether $\tilde{g}_{rr} \gtrless 1$ (see \eref{embfunc}). The
embedding diagram is described by the embedding function $z(r)$,
such that the surface $z(r)$ has the same intrinsic geometry as
the original spacetime slice:
\begin{equation}\label{embfunc}
z(r)=\int\sqrt{\pm(\tilde{g}_{rr}-1)}dr
\end{equation}
where the $\pm$ signs are for Euclidean and Minkowski embeddings
respectively. Timelike hypersurfaces $(ds^2<0)$, which imply
$\tilde{g}_{rr}<0$ everywhere, are always embedded in Minkowski
space, whereas certain spacelike hypersurfaces $(ds^2>0)$ are
embedded in Euclidean space and others in Minkowski space.
\par
\begin{figure}[ht]
   \centering
   \includegraphics[height=10cm,clip=true]{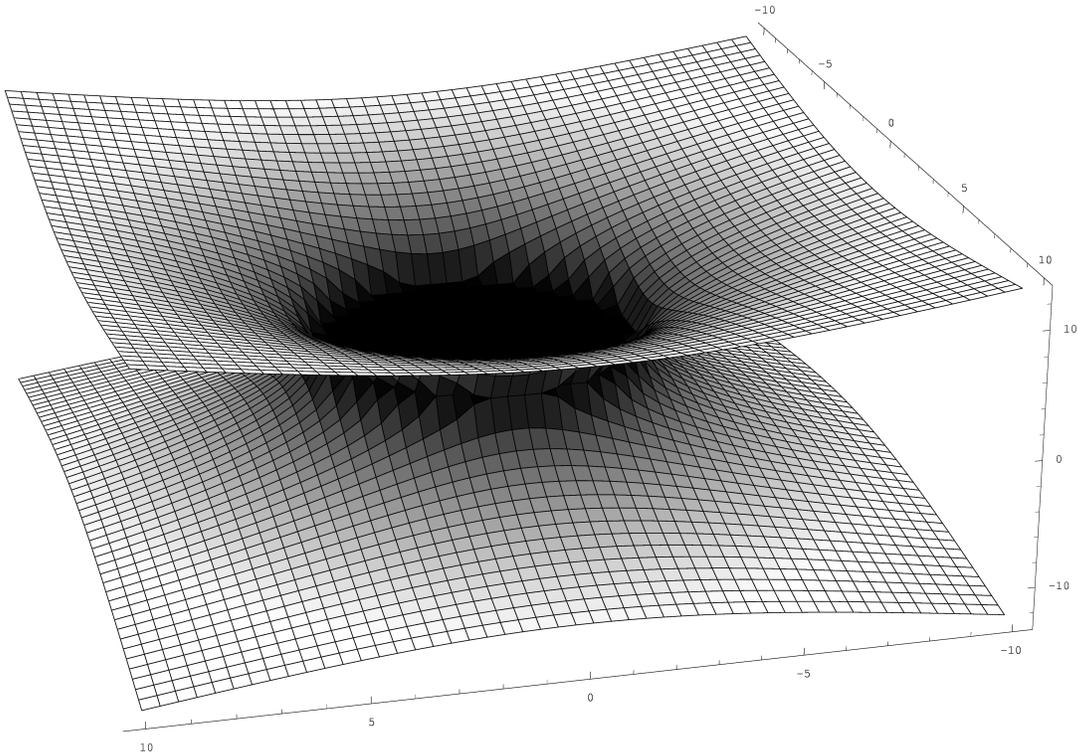}
   \caption{An embedding diagram in Euclidean space of a \SC
spacetime hypersurface described by a horizontal line in the \KS
coordinates. This surface displays a spacelike wormhole. The
diagram shows the vertical embedding lift $z$ as a function of the
$r,\varphi$ coordinates (the $x-y$ plane).}
   \label{fig:sch-hor}
\end{figure}
\begin{figure}[ht]
   \centering
   \includegraphics[width=13cm,clip=true]{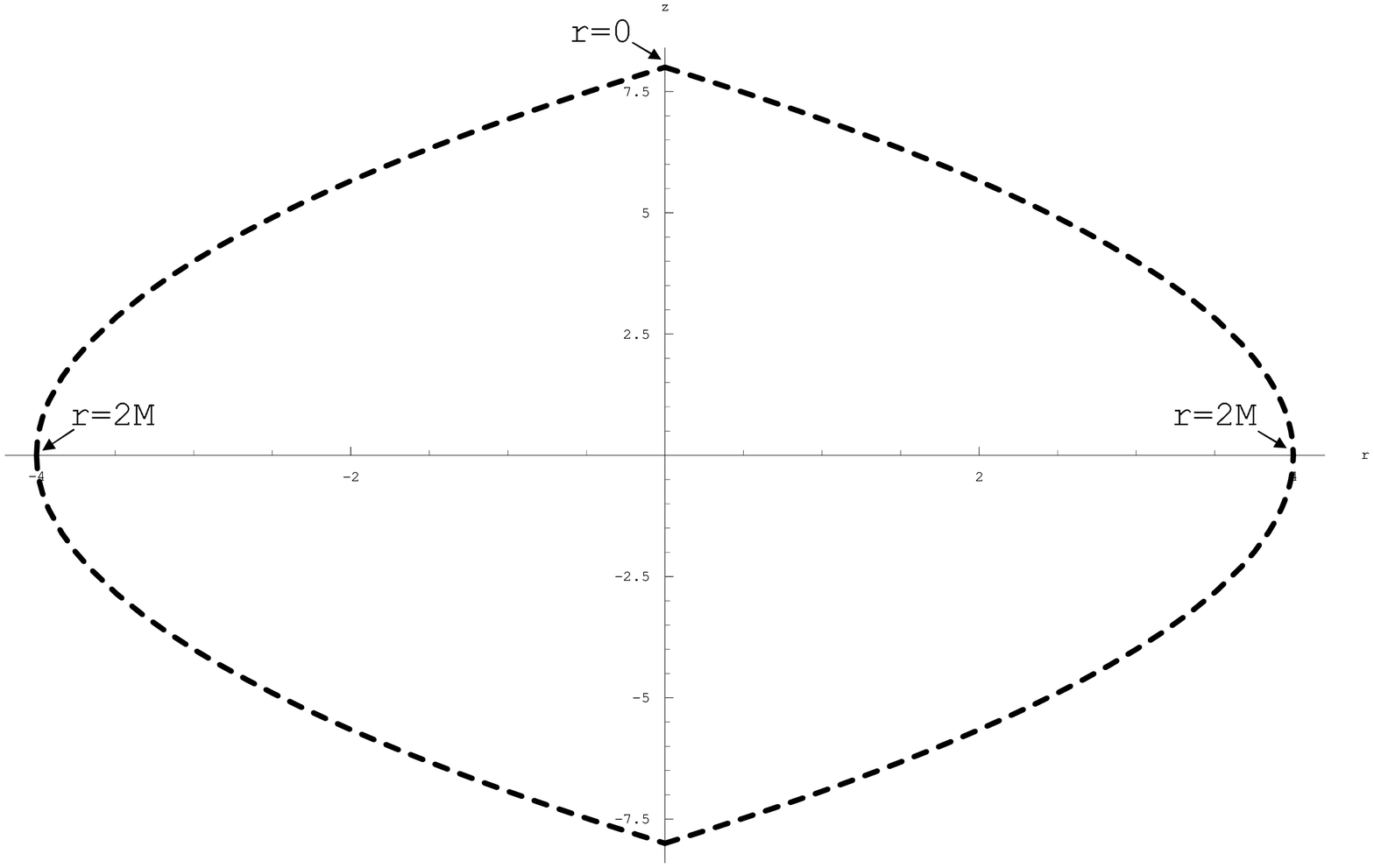}
   \caption{An embedding diagram in Minkowski space of a timelike
\SC spacetime hypersurface described by a vertical line in the \KS
coordinates. Here we discard of the symmetric $\varphi$ coordinate
and display the embedding lift as a function of $r$ alone. The
left half of the diagram does not refer of course to negative
$r$'s - it is merely a rotation of 180$^\circ$ that is shown in
order to ease the visualization of the entire surface. This
embedding is denoted by a dashed line to indicate the
background Minkowski space.}
   \label{fig:sch-ver}
\end{figure}

In 1916 Flamm \cite{Flamm} had already obtained the
well known \SC ``wormhole" with an analytic embedding function of
$z=[8M(r-2M)]^{\frac{1}{2}}$. This describes a hypersurface of
$t=const$ outside the event horizon, and similar embeddings
correspond to horizontal lines, $v=const$, in the \KS coordinates.
An embedding diagram of such a hypersurface is displayed in figure
\ref{fig:sch-hor}.
The less familiar embedding of the $t=const$ hypersurface inside
the \SC event horizon $(r<2M)$ is presented in figure
\ref{fig:sch-ver}. This is a timelike hypersurface (described by a
vertical line, $u=const$, in the \KS coordinates), and its
embedding is in Minkowski space.
Since our spacetime slices always have an axial symmetry, we will
use in the following only 2-dimensional diagrams such as figure
\ref{fig:sch-ver} to describe our surfaces. We adopt here a
convention that when the embedding space is Minkowski the
diagram is described by a dashed line, while it is a solid line in
Euclidean space.

\section{The coordinate systems}\label{COOsec}
The familiar RN metric is described in Schwarzschild-like
coordinates as:
\begin{equation}\label{origincoo}
ds^2=-\left(1-\frac{2M}{r}+
\frac{Q^2}{r^2}\right)dt^2+\left(1-\frac{2M}{r}+\frac{Q^2}{r^2}\right)^{-1}dr^2+
r^2(d\theta^2+sin^2\theta d\varphi^2)
\end{equation}
where the parameters $M$ and $Q$ describe the mass and electric
charge. We consider here $M>|Q|$ when a black hole exists. In this
case we have two horizons:
\begin{equation}
r_\pm=M\pm\sqrt{M^2-Q^2}
\end{equation}
We get rid of the coordinate singularities on $r_\pm$ by moving to
a Kruskal-Szekeres-like coordinate system:
\begin{equation}
ds^2=f(r)(-d\psi^2+d\xi^2)+r^2(d\theta^2+sin^2\theta d\varphi^2)
\end{equation}
where $f(r)>0$. In fact we use two separate coordinate systems -
coordinate system A that is analytic at $\Rp$ but singular at
$\Rm$ and coordinate system B that is analytic at $\Rm$ but
singular at $\Rp$. The coordinate transformations from $r,t$ to
$\xi,\psi$ will take a different form in the three regions of
spacetime:
\begin{align*}
I:\;\;\;\;\;\;\;\;\;\;\;r>\Rp \\
II:\;\Rm<r<\Rp \\
III:\;\;\;\;\;\;\;\;r<\Rm
\end{align*}
We will use system A to describe regions $I$ and $II$ and system B
to describe regions $II$ and $III$. The transformations to
coordinate systems A,B in regions $I,II,III$ are:
\begin{subequations}\label{transformations}
\begin{align}
\left\{%
\begin{array}{c}
 \xi_{A,I} \\
  \xi_{A,II} \\
  \xi_{B,II} \\
  \xi_{B,III} \\
\end{array}
\right\}%
=\left\{\begin{array}{c}
 C(r) \\
  C(r) \\
  C(r)^{-1} \\
  -C(r)^{-1} \\
\end{array}\right\}
\left\{\begin{array}{c}
 \cosh[F(t)] \\
  \sinh[F(t)] \\
   \sinh[F(t)]\\
  \cosh[F(t)]\\
\end{array}\right\}
\\
\left\{%
\begin{array}{c}
 \psi_{A,I} \\
  \psi_{A,II} \\
  \psi_{B,II} \\
  \psi_{B,III} \\
\end{array}
\right\}%
=\left\{\begin{array}{c}
 C(r) \\
  C(r) \\
  -C(r)^{-1} \\
  C(r)^{-1} \\
\end{array}\right\}
\left\{\begin{array}{c}
 \sinh[F(t)] \\
  \cosh[F(t)] \\
   \cosh[F(t)]\\
  \sinh[F(t)]\\
\end{array}\right\}
\end{align}
\end{subequations}
where
\begin{equation}
F(t)\equiv\frac{\sqrt{M^2-Q^2}}{\Rp^2}t
\end{equation}
and
\begin{equation}
C(r)\equiv\left|\frac{r-\Rp}{2M}\right|^{\frac{1}{2}}\left|\frac{r-\Rm}{2M}
\right|^{-\frac{\Rm^2}{2\Rp^2}}\exp\left[\frac{\sqrt{M^2-Q^2}}{\Rp^2}r\right].
\end{equation}
$C(r)$ is an analytic function in each region of spacetime, and $r$ is expressed implicitly in terms of the new coordinates using:
\begin{equation}
\xi^2-\psi^2=\pm C(r)^{\pm2}
\end{equation}
where the appropriate $\pm$ signs for each region and coordinate system is determined by \eref{transformations}. We use in the following lower indexes to indicate the coordinate system and region to which we refer.
\par
\begin{figure}[ht]
   \centering
   \includegraphics[height=9cm,width=9cm,clip=true]{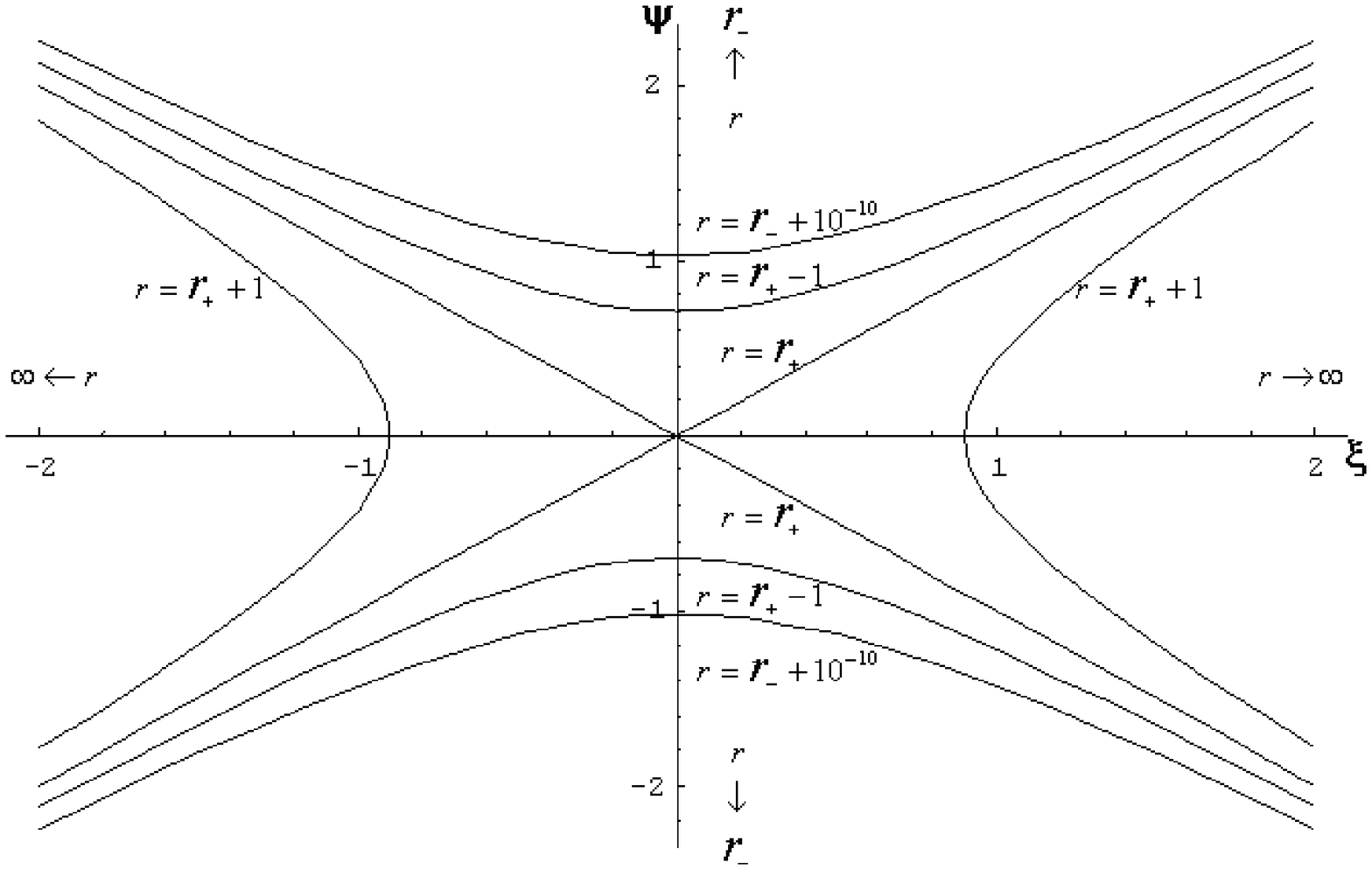}
   \caption{Coordinate system A, analytic around the $\Rp$ horizon.
The relation between the original radial coordinate and the new
coordinates is displayed for an example of a black hole
characterized by $M=2,\;Q=1$. $r=const$ hypersurfaces appear here
as hyperbolae with the asymptotes $\psi=\pm\xi$. These asymptotes
describe the surfaces $r=\Rp$, while the surfaces $r=\Rm$, which
are singular in the current coordinates, are located at the upper
and lower infinity of this diagram (where
$\xi^2-\psi^2=-\infty$).}
   \label{fig:coosysA}
\end{figure}

Figure \ref{fig:coosysA} depicts constant $r$ lines in the
$\xi,\psi$ plane of coordinate system A.
The transformations presented in \eref{transformations} are valid
for one set of system A regions -
the upper right half of figure \ref{fig:coosysA} corresponding to
$\xi>|\psi|$ or $\psi>|\xi|$. A parallel set of regions with
identical geometries is obtained by similar transformations
$(\xi_A\rightarrow-\xi_A,\;\psi_A\rightarrow-\psi_A)$.
\par
\begin{figure}[ht]
   \centering
   \includegraphics[height=9cm,width=9cm,clip=true]{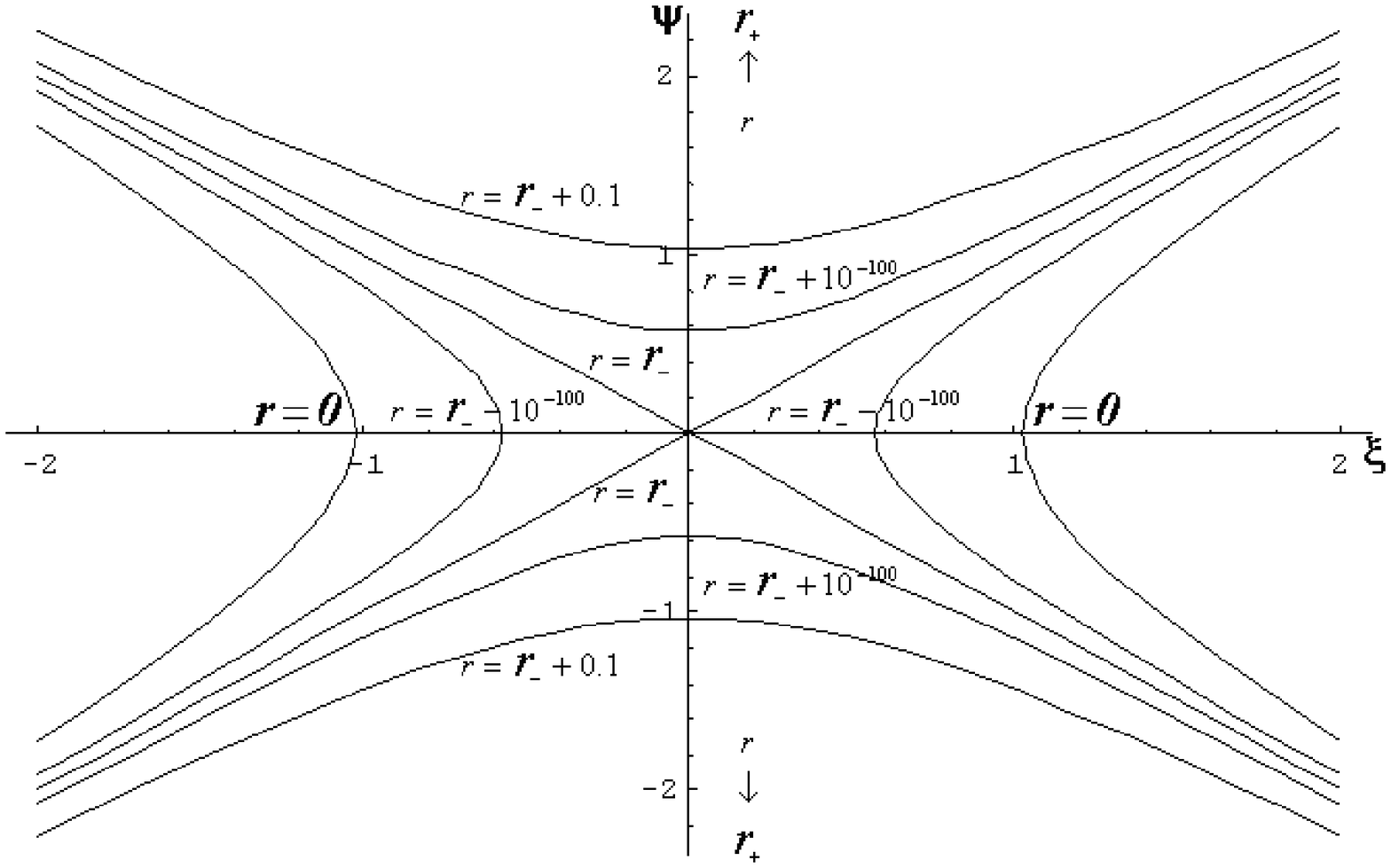}
   \caption{Coordinate system B, analytic around the $\Rm$ horizon.
   The relation between $r$ and the new coordinates is displayed for an example of $M=2,\;Q=1$. $r=const$ hypersurfaces appear here as hyperbolae with the asymptotes $\psi=\pm\xi$. These asymptotes describe the surfaces $r=\Rm$, while the surfaces $r=\Rp$, which are singular in the current coordinates, are located at the upper and lower infinity of this diagram.}
   \label{fig:coosysB}
\end{figure}

Figure \ref{fig:coosysB} shows constant $r$ lines in the
$\xi_B,\psi_B$ plane.
The system B transformations in \eref{transformations} describe
the lower left half of figure \ref{fig:coosysB}, and again the
parallel regions are similarly obtained. The two separate
coordinate systems connect of course as $\psi$ goes to
$\pm\infty$, demonstrating the known structure of the RN
spacetime, which is in essence an infinite chain of parallel
asymptotically flat universes connected by wormholes.
\par
We consider embedding diagrams of spacelike and timelike hypersurfaces
that are described by horizontal and vertical lines in the
$\xi,\psi$ coordinate systems. Horizontal lines mean $d\psi=0$,
which implies in coordinate system A:
\begin{equation}
\left\{%
\begin{array}{c}
 dt_{I} \\
 dt_{II} \\
\end{array}
\right\}%
=
- \left\{%
\begin{array}{c}
 \tanh[F(t)] \\
  \coth[F(t)] \\
\end{array}
\right\}%
\left(1-\frac{2M}{r}+\frac{Q^2}{r^2}\right)^{-1}dr
\end{equation}
Using \eref{embfunc}, the embedding function of such a hypersurface
in Euclidean $(+)$ or Minkowski $(-)$ space is given by:
\begin{equation}\label{horembfunc}
\left\{%
\begin{array}{c}
 z(r)_{I} \\
 z(r)_{II} \\
\end{array}
\right\}%
=\int\sqrt{\pm\left(\frac{2M}{r}-\frac{Q^2}{r^2}-
\left\{%
\begin{array}{c}
 \tanh^2[F(t)]\\
 \coth^2[F(t)] \\
\end{array}
\right\}%
\right)\left(1-\frac{2M}{r}+\frac{Q^2}{r^2}\right)^{-1}}dr
\end{equation}
Repeating this procedure for horizontal lines in region $II$ of
coordinate system B we get the same embedding formula as the one
for coordinate system A in \eref{horembfunc}. The embedding
function of horizontal lines in region $III$ of coordinate system
B is identical to the one for region $I$ in \eref{horembfunc}. The
difference is that the parameter $t$ is assigned different values
than before, as it is determined by solving the equation
$\psi_B=const$ rather than $\psi_A=const$.
\par
We can similarly arrive at the expressions for embedding surfaces
of vertical lines $(d\xi=0)$. Such a vertical line is partly
described by coordinate system A and partly described by
coordinate system B. In order to get the entire embedded surface
(which passes through regions $I$, $II$ and $III$) we need to
connect continuously and smoothly the embedding curves of the
two coordinate systems. Note that due to the different
coordinate transformations, straight lines of $\xi=const$ in
coordinate system A do not necessarily connect with straight lines
of the same $\xi=const$ in coordinate system B.
Once we choose the connection point between $\Rm$ and
$\Rp$, we are required to match
the same spacetime event in both coordinate systems, determining
specific sets of vertical lines. Regardless of this
arbitrariness, the embedding diagrams show the qualitative features
of timelike and spacelike hypersurfaces, and the arbitrary details
of the specific hypersurfaces are not important. The embedding functions
we get in Minkowski space (timelike surfaces can never be embedded
in Euclidean space) are:
\begin{equation}
\left\{%
\begin{array}{c}
 z(r)_{I} \\
 z(r)_{II} \\
 z(r)_{III} \\
\end{array}
\right\}%
=\int\sqrt{-\left(\frac{2M}{r}-\frac{Q^2}{r^2}-
\left\{%
\begin{array}{c}
 \coth^2[F(t)]\\
  \tanh^2[F(t)] \\
 \coth^2[F(t)] \\
\end{array}
\right\}%
\right)\left(1-\frac{2M}{r}+\frac{Q^2}{r^2}\right)^{-1}}dr
\end{equation}
where $t$ is determined by coordinate system A's transformation for region $I$ and by
coordinate system B's transformation for region $III$. For region $II$ $t$ can be
determined by either systems' transformations, depending on our choice of the vertical
lines' connection point.

\section{The embedding diagrams of RN hypersurfaces}\label{RNEMBsec}
\begin{figure}[ht]
   \centering
   \includegraphics[height=15cm,clip=true]{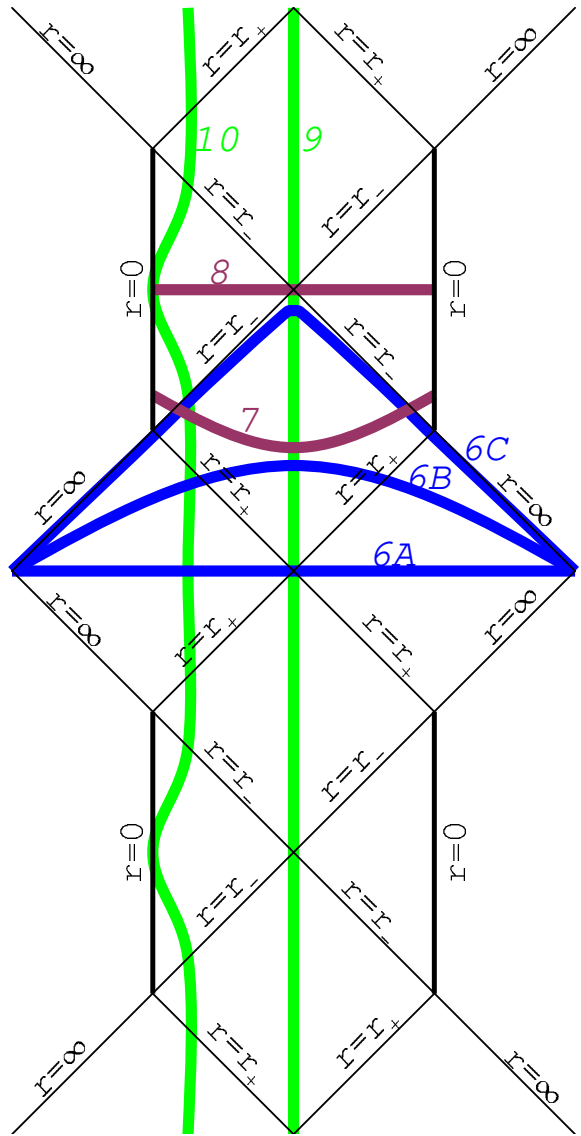}
   \caption{A Penrose diagram of the RN spacetime.
   The thick lines describe the hypersurfaces, whose embedding diagrams
   we display. The numbering corresponds to the embedding figures' numbers. While the straight lines \textit{(6A,8,9)} provide an accurate description of the embedded surfaces, the other lines are schematic.}
   \label{fig:Penrose}
\end{figure}

Using the tools developed in the previous section we turn now to
the embedding diagrams of different hypersurfaces in the RN
spacetime. Figure \ref{fig:Penrose} depicts a Penrose diagram of
the RN spacetime (as shown in \cite{Carter}).
We sketch on this diagram lines corresponding to the hypersurfaces
for which we present embedding diagrams in this section. We use
as an example throughout the discussion here the case $M=2,\;Q=1$.
\par
We begin with spacelike hypersurfaces presented as horizontal lines
in coordinate system A. The line element of the $\psi=0$ surface
is the same as any $t=const$ surface $(r>\Rp)$ in the original
$r,t$ coordinates. We expect this surface to be similar to that of
$t=const,\;r>2M$ in the \SC spacetime, as both cases describe a
spacelike wormhole connecting at the event horizon two
asymptotically flat regions. The spacetime geometry is independent
of the $t$ coordinate, but $t$ becomes a spacelike coordinate when
crossing the event horizon. This means that for an observer
looking at the black hole from outside the spacetime appears
static. To explore the geometry seen by an observer who crosses
the event horizon and journeys into the black hole we advance with
the $\psi$ coordinate, which is a genuine timelike coordinate in
all regions, and inspect the evolution of the spacelike
hypersurfaces with $\psi=const>0$.
\par
\begin{figure}[ht]
   \centering
   \includegraphics[width=13cm,clip=true]{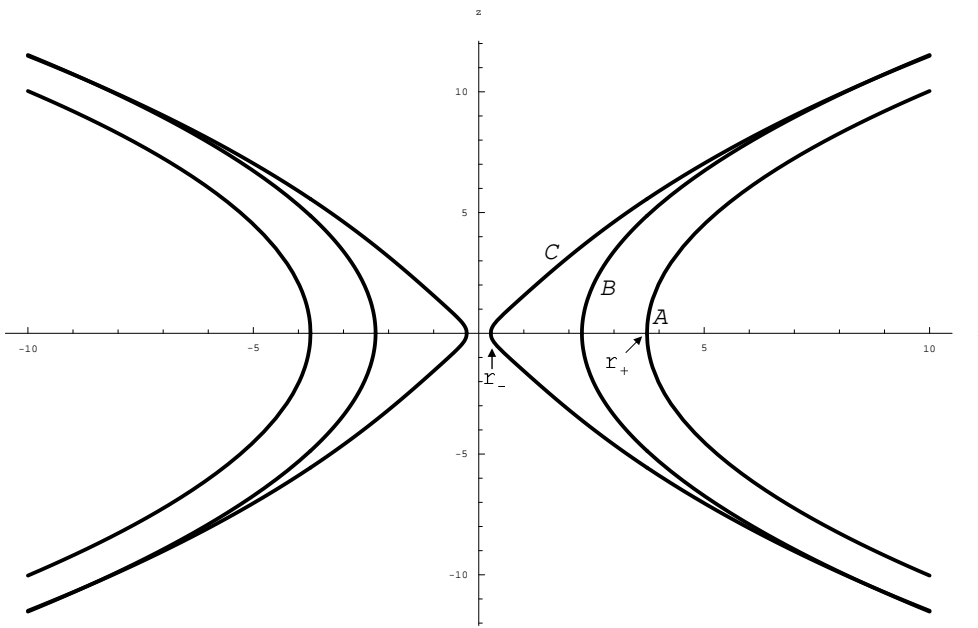}
   \caption{Embedding diagrams in Euclidean space of spacelike RN
   hypersurfaces described by horizontal lines in coordinate system A. These embeddings display the evolving nature of the spacelike wormhole. Curve \textit{A} describes the $\psi=0$ surface where the wormhole is widest. Curve \textit{B} describes the $\psi=0.8$ surface, and curve \textit{C} belongs to the $\psi=1$ surface, which nearly describes the wormhole of smallest circumference.}
   \label{fig:emb123}
\end{figure}

Figure \ref{fig:emb123} depicts the embedding diagrams of the
$\psi_A=0$, $\psi_A=0.8$ and $\psi_A=1$ hypersurfaces.
The geometry of the different surfaces varies. This demonstrates
the fact that the RN spacetime is actually dynamic. As we
advance in $\psi_A$ the wormhole shrinks. This phenomena also occurs
when advancing along the timelike \KS coordinate in the \SC
spacetime. While $\Rm$ is reached only as $\psi_A \rightarrow
\infty$, the surface described by $\psi_A=1$
can be treated with our working accuracy as if it is tangent to
the $r=\Rm$ hyperbola (see figure \ref{fig:coosysA}). The wormhole
with a throat radius of $\Rm$ is the narrowest wormhole that the
time evolution in coordinate system A allows. This is contrary to
the \SC case, where the wormhole continues to shrink until it
closes when reaching the singularity.
\par
\begin{figure}[ht]
   \centering
   \includegraphics[width=13cm,clip=true]{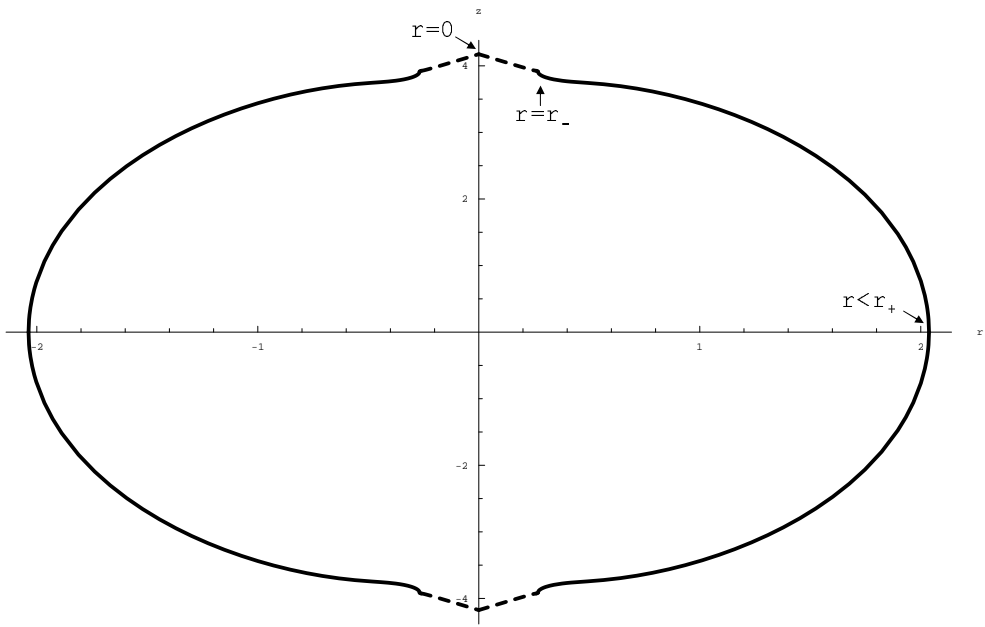}
   \caption{The embedding diagram of the
$\psi_B=-1.19$ hypersurface. For brevity the figure depicts
two different surfaces that describe the embeddings of parts of
this hypersurface in different background spaces. The portion
described by dashed lines is embedded in Minkowski space and the
portion described by solid lines is embedded in Euclidean space.
We can see an inflection point at $r=\Rm$. Note that the Minkowski
portion starts at $r<\Rm$.}
   \label{fig:emb4}
\end{figure}
\begin{figure}[ht]
   \centering
   \includegraphics[width=13cm,clip=true]{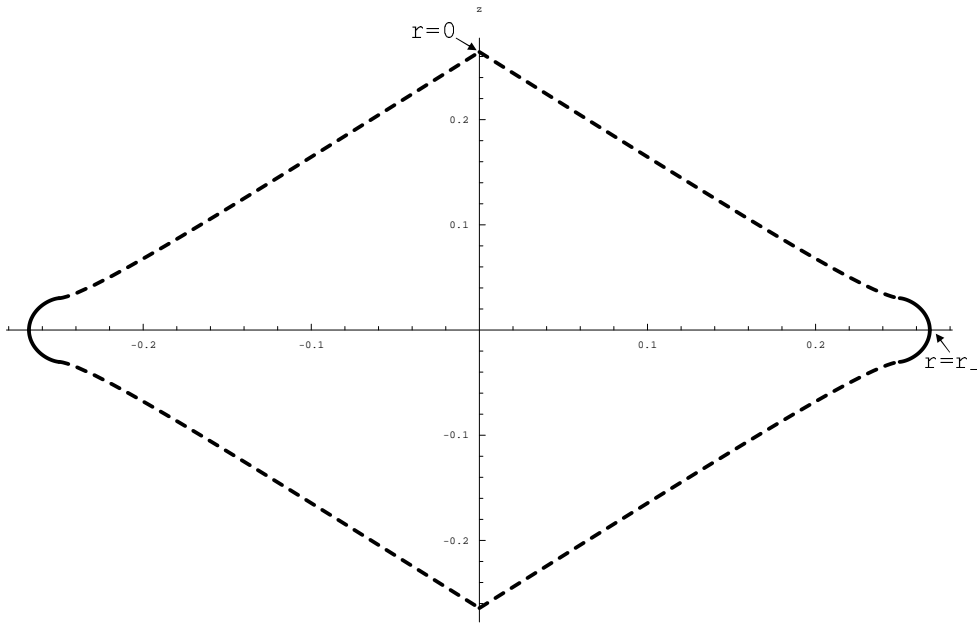}
   \caption{The embedding diagram of the $\psi_B=0$ hypersurface.
For brevity, here again, the figure depicts
two different surfaces that describe the embeddings of parts of
this hypersurface in different background spaces.}
   \label{fig:emb5}
\end{figure}

To explore further the RN spacetime we must transform to
coordinate system B. Horizontal lines in this coordinate system
describe spacelike hypersurfaces that begin at the $r=0$ singularity
and increase in radial distance until connecting with the parallel
region after crossing the $\Rm$ horizon. Therefore the embedding
diagrams of these hypersurfaces will be of a different nature. In
the proximity of the $r=0$ singularity we can never embed any
surface in Euclidean space. The RN line element \eref{origincoo}
has $g_{rr}\rightarrow0$ and $g_{tt}\rightarrow-\infty$ as
$r\rightarrow0$. Every path that approaches $r=0$ will have at a
sufficiently small radius $\tilde{g}_{rr}<1$. Therefore  around
the singularity we must embed the following surfaces in Minkowski
space. At a certain distance the embedding becomes Euclidean. As
we examine the evolution of spacelike surfaces seen by a traveler
that enters the black hole, we enter coordinate system B from the
bottom and draw in figure \ref{fig:emb4} the embedding diagram of
the surface described by $\psi_B=-1.19$.
The connection between the parallel regions $(z=0)$ occurs now
when increasing the radial distance (up to $r=2<\Rp$ in the
example above). Figure \ref{fig:coosysB} shows that the two inner
parallel regions are "closest" in the surface of $\psi_B=0$, where
they touch at $r=\Rm$. The embedding of this surface is displayed
in figure \ref{fig:emb5}.
\par
It is interesting to inspect the resemblance between the geometry
of the horizontal lines in coordinate system B and the geometry of
the vertical lines in the \KS coordinates of the \SC spacetime. We
can imagine a resemblance because the \SC spacetime's vertical
lines also emerge from the $r=0$ singularity and connect with a
parallel region after crossing the horizon. However, these
vertical lines describe timelike hypersurfaces and so their
embeddings will all be in Minkowski spaces. Comparing figures
\ref{fig:emb4},\ref{fig:emb5} with figure \ref{fig:sch-ver}, we
indeed see some resemblance. The region near the singularity in
the RN spacetime's horizontal lines has a similar general shape to
the \SC spacetime's vertical lines on their way from the
singularity to the parallel region. The region in RN with $r>\Rm$
continues towards the parallel region like the \SC surfaces, but
this region is embedded in Euclidean space while the \SC
surfaces continue in Minkowski space.
\par
\begin{figure}[ht]
   \centering
   \includegraphics[width=13cm,clip=true]{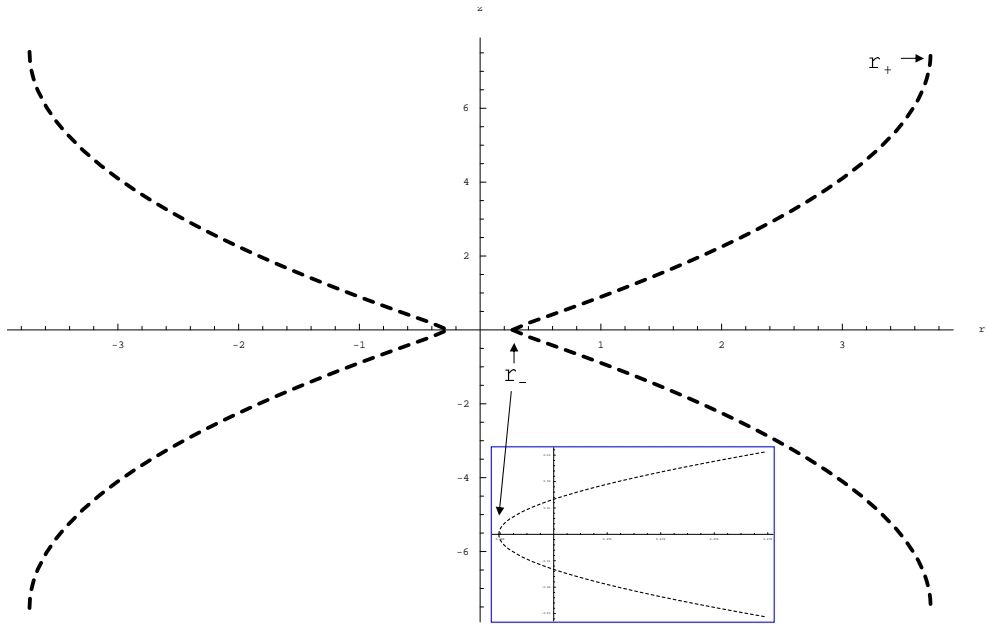}
   \caption{The Minkowski embedding of the $\xi=0$ hypersurface. The diagram shows just one of the periodic segments of the infinite surface. The enlargement demonstrates the smoothness of the embedding at the $\Rm$ wormhole throat.}
   \label{fig:emb6}
\end{figure}

We turn now to inspect timelike hypersurfaces of the RN spacetime
described by vertical lines in our coordinate systems. The $\xi=0$
line passes in the two coordinate systems only in regions $II$
between $r=\Rm$ and $r=\Rp$. The embedding diagram of this surface
(in Minkowski space of course) is displayed in figure
\ref{fig:emb6}.
We can see in the figure a connection (the timelike wormhole
presented in coordinate system B) between parallel regions at
$r=\Rm$. Additionally, we can see in coordinate system A a
timelike wormhole connecting parallel regions at $r=\Rp$, and this
should be visible in the embedding by connecting the upper and
lower ends of figure \ref{fig:emb6} with identical diagrams. The
complete surface is achieved by connecting an infinite chain of
such diagrams. When passing through the $\Rm$ horizon to a
parallel region there appears to be a cusp in the embedding. This
is not a consequence of performing a wrong connection between the
coordinate systems, since the entire calculation here was done in
a single coordinate system. However, a close inspection of the
embedding function reveals that its slope is in fact continuous at
$r=\Rm$. When approaching the wormhole throat at $\Rm$, $z'$
diverges, as is required for a smooth connection (see the inset
in figure \ref{fig:emb6}).
\par
\begin{figure}[ht]
   \centering
   \includegraphics[width=13cm,clip=true]{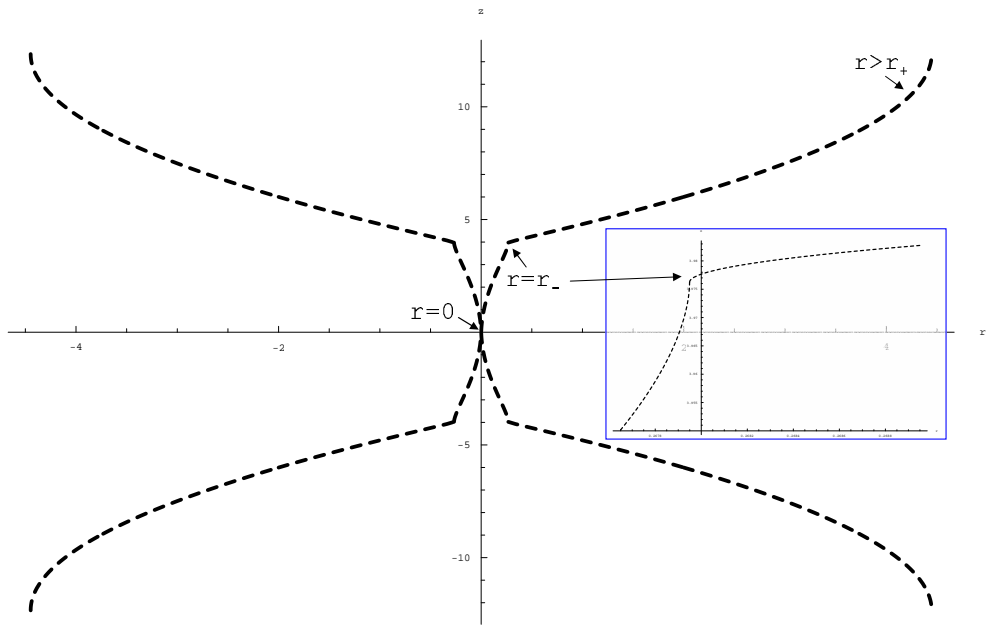}
   \caption{The Minkowski embedding of a timelike hypersurface
   described by a vertical line that is tangent to the $r=0$ surface.}
   \label{fig:emb7}
\end{figure}

We now move the vertical line to the side (either positive or
negative $\xi$) and examine a hypersurface that touches the
singularity at $r=0$ (see figure \ref{fig:emb7}).
This timelike surface passes regions $I$, $II$ and $III$. It
begins by touching the $r=0$ surface, where the wormhole throat is
closed. Its radius increases and it crosses the $\Rm$ horizon
(where the embedding seems again to break). It continues past the $\Rp$ horizon and reaches a
maximal radius where it connects to a new parallel region. The
complete surface is, of course, described by an infinite chain of
identical diagrams. It should be noted that at $r=\Rm$ we
were careful not to switch between coordinate systems (we have a
freedom to connect the vertical lines of the two coordinate
systems somewhere between $r=\Rm$ and $r=\Rp$). The surface is
smooth at the point of the coordinate systems' connection
($r=(\Rm+\Rp)/2=M=2$ in this example). Again, the surface is in fact also smooth at $r=\Rm$, as a detailed inspection of this area shows.
\par
The timelike wormhole around the singularity is widest (the
throat has a radius of $\Rm$) for the hypersurface described by
$\xi=0$. This is similar to the surface described by the
horizontal line $v=0$ in the \KS \SC coordinates (described in
figure \ref{fig:sch-hor}), where the spacelike wormhole is widest
(with a throat radius of 2M). When we move in the \KS
coordinates towards horizontal lines of larger $v$, the wormhole
shrinks until it closes with zero circumference at the $r=0$
singularity. This resembles our case, where moving the
vertical lines towards larger $\xi$ caused the wormhole to shrink
until it eventually closed at the hypersurface tangent to $r=0$.
A resemblance between the known \SC embedding
diagrams and the RN embedding diagrams is indeed apparent.
Notice again the significant difference that the relevant \SC
surfaces are spacelike and embedded in Euclidean space, whereas
our current surfaces are timelike and embedded in Minkowski space.
The character of the $\Rm$ horizon is different than that of the
\SC event horizon, and this is apparent in the diagrams. In
addition, the \SC spacetime horizontal lines continue towards
radial infinity, whereas our vertical lines cross another horizon
at $\Rp$ and then reach another wormhole and contract towards
$\Rm$. Thus, the resemblance between the \SC horizontal lines and
the RN vertical lines is limited to the region of coordinate
system B, describing the area around the singularity and the $\Rm$
horizon in the RN spacetime. The portions of the vertical lines
that reside in coordinate system A (around the $\Rp$ horizon) are
in fact similar to vertical lines in the \SC spacetime, as is
evident by a comparison to figure \ref{fig:sch-ver}. This is not
surprising due to the similarity of coordinate system A to the \KS
coordinates.

\section{Summary}\label{SUMsec}
We developed tools and displayed embedding diagrams of spacelike and timelike hypersurfaces of the RN spacetime. The hypersurfaces described provide a typical example of spacelike and timelike slices of the RN spacetime and demonstrate the evolution of RN surfaces. The visualization presented for these slices provides an alternative way to understand intuitively the geometry of the RN spacetime and its properties.
\par
The RN geometry is similar to the \SC geometry in the region
external to the event horizon $(r>\Rp)$.
In accordance with this the embedding diagrams of hypersurfaces
described in the coordinate system that crosses $r=\Rp$ were
similar to the embedding diagrams of the analogous \SC
hypersurfaces. When we advance in time to different RN spacelike
slices described by horizontal lines in this coordinate
system, the wormhole shrinks, as is the case when the \SC
spacetime evolves in the time coordinate of Kruskal-Szekeres.
However, as expected, when we continue to
advance in time and reach the hypersurfaces described in the
coordinate system that crosses $r=\Rm$, the similarity to \SC
stops. The $\Rm$ horizon has no analogy in the \SC spacetime,
and the RN horizontal lines that intersect $r=\Rm$ exhibit a
different behavior. These hypersurfaces hit on both sides the
$r=0$ singularity, and their geometry near the singularity
resembles the geometry of timelike \SC hypersurfaces,
presented as vertical \KS lines.
\par
Finally, vertical lines in the RN coordinates describe a path
through a wormhole passing beside the singularity. The geometry of
the respective surfaces has some features resembling the geometry
of the surfaces describing the \SC wormhole, which are presented
there as horizontal lines. The fundamental difference is that the
\SC wormhole is a spacelike path that can not be embarked upon,
whereas the RN wormhole is an actual timelike path.
This is reflected in the evolution of the spacelike hypersurfaces.
When we advance in time the \SC wormhole shrinks and closes.
However, progressing with
the RN spacelike surfaces does not cause the wormhole to
close, as apparent from our embedding diagrams, and we are not
forced to be drained into the singularity. Instead one may choose to
take a path like the ones described by our timelike embedding diagrams,
and pass through the timelike wormhole
to a parallel region and to infinite new universes.

\ack
We thank Nathalie Deruelle, Joseph Katz and an anonymous referee for helpful remarks. This research was supported by the ISF center of excellence for High Energy Astrophysics and by the Schwartzmann University Chair (TP).

\end{document}